\documentclass[conf]{new-aiaa}
\usepackage[utf8]{inputenc}
\usepackage{graphicx}
\usepackage{amsmath}
\usepackage[version=4]{mhchem}
\usepackage{siunitx}
\usepackage{longtable,tabularx}
\usepackage{float}
\usepackage{caption}
\DeclareCaptionLabelFormat{boldlabel}{\textbf{#1~#2}}
\DeclareCaptionFont{myitalic}{\normalfont\itshape}
\title{Kirigami Film Reflector for Deployable Space Antennas}

\author{Gulzhan Aldan\footnote{Graduate Student, Department of Mechanical Engineering and Applied Mechanics, AIAA Student Member}, Henry Love\footnote{Graduate Student, Department of Electrical and Systems Engineering}, Matthew Campbell\footnote{Research Assistant Professor, Department of Mechanical Engineering and Applied Mechanics}, Firooz Aflatouni\footnote{Professor, Department of Electrical and Systems Engineering}, Igor Bargatin\footnote{Associate Professor, Department of Mechanical Engineering and Applied Mechanics, AIAA Member}\footnote{Corresponding Author. Email: bargatin@seas.upenn.edu}}
\affil{University of Pennsylvania, Philadelphia, Pennsylvania, 19104, United States}

\begin{document}

\bibliographystyle{ieeetr}

\maketitle

\begin{abstract}

We propose a low-pretension reflective kirigami film as a material for the reflective surfaces of large deployable space reflector antennas with an operating frequency around 10~GHz. The kirigami cut pattern is based on the well-known rotating squares pattern but is augmented with diagonal cuts to enhance stretchability and allow control over the effective Poisson's ratio. Using finite element simulations, we analyzed how the geometric parameters of this pattern affected the reflectance of the film and the pretension required to resist thermal deformations. Tensile testing of selected designs, which are approximately half the weight of traditional metallic meshes, demonstrated a substantial reduction in the needed pretension to $\sim$0.5~N/m and as low as $\sim$0.1~N/m. Such low pretension represents an order-of-magnitude improvement over traditional metallic mesh reflectors and could enable the use of lighter antenna trusses. Free-space reflectance measurements also show that these perforated films can maintain power reflectance exceeding 90\% at 10~GHz under the strains expected in the deployed configuration.

\end{abstract}

\section{Introduction}

\lettrine{S}{ince} the dawn of space exploration, the launch of large space structures has been fundamentally limited by the Tsiolkovsky equation and the size of rocket fairings. These constraints on payload mass, volume, and maximum linear dimensions have driven the development of ultra-lightweight deployable structures that can be compactly packed on Earth and reliably unfold in orbit \cite{Schenk2014, miura2020forms}. Such structures are particularly important for realizing large space-based reflector antennas \cite{Tibert2002}, which enable high angular resolution and data transmission over long distances \cite{Hansen1981}. Today, the largest commercial antennas are up to 22~m in diameter \cite{northrop2017} and use the AstroMesh architecture where a reflective surface is a metallic mesh pretensioned over a deployable circular truss \cite{Thomson1999}. However, future radio frequency (RF) reflectors for Earth observation, satellite communication, and astrophysical measurements will require apertures hundreds of meters wide \cite{POWELL1978}, which are not yet achievable with existing deployment methods.

A main challenge in constructing large space antennas is maximizing the reflective surface area while maintaining the mass and stowed volume of the entire structure, including the truss, within the constraints of available launch vehicles. One possible solution is to forego the deployment of the preassembled stowable truss in favor of disassembled tubular truss components, which can be stowed more efficiently and autonomously assembled in space using robotic systems \cite{Lee2016, Suh2024, dassanayake2025numerical}. As a proof of concept, recent studies have demonstrated a robotic truss builder capable of retrieving individual components from the storage space and constructing a complete lab-scale prototype \cite{Suh2024}. This in-space assembly approach could enable the deployment of trusses for reflector antennas with diameters of up to 200~m \cite{Dassanayake2024}.

Minimizing the mass and stowed volume of the perimeter truss also requires reducing the pretension needed in the reflective mesh. Traditional designs typically use gold-plated molybdenum mesh wires with diameters of 30--45~\textmu m and an areal density of $\sim$25~g/m\textsuperscript{2} \cite{Tibert2002, Imbriale1991, Decius2023, hptex}. However, to ensure high reflectance, stable low-resistance electrical contact needs to be maintained between the wires, which requires constant pretension of about 5~N/m (ranging from 2 to 100~N/m) \cite{Dassanayake2024, Hedgepeth1983, Thomson1999, Tibert2002, santiago, Ozawa2023}. In the AstroMesh architecture, the reflective mesh is pretensioned against a cable net attached to the perimeter truss, which in turn induces compressive loads on the truss elements. To withstand compression and prevent buckling of the structure, the tubular truss members must be sufficiently thick or large in diameter. Lowering the required mesh pretension could reduce the compressive forces acting on the truss and, as a result, allow thinner and lighter truss components and a smaller stowed volume.

In this work, we propose a low-pretension perforated aluminized polyimide film as an alternative reflective surface for large AstroMesh antennas. Polyimide is a thermally and chemically stable material widely used in electronics \cite{Ji2019} and space applications \cite{Gouzman2019}, including solar sail missions such as IKAROS \cite{ikaros} and NanoSail-D2 \cite{katan2012nasa}.  Since the operational temperature of the aluminized polyimide ranges from $-250$~\textdegree{}C to 290~\textdegree{}C \cite{Sheldahl2020}, it is able to withstand thermal cycling from $-200$~\textdegree{}C to 100~\textdegree{}C caused by periodic exposure to sunlight and shadow in orbit. Additionally, the thinnest commercially available polyimide films coated with 100~nm aluminum have an areal density of 11~g/m\textsuperscript{2}, which is about half that of a gold-plated molybdenum mesh \cite{Sheldahl2020}. Such a contiguous film-based reflector eliminates the need for the high pretension required to maintain electrical contact between the wires, and, therefore, enables a lighter truss design. 

While the use of polyimide eliminates the need for high pretension, low-level pretension is still required to resist thermal deformations caused by differential thermal expansion and temperature gradients during thermal cycling. Polyimide has a thermal expansion coefficient of $\sim$32~ppm/K \cite{DuPont2022} that results in thermal strain variations of slightly less than 1\% under typical orbital temperature swings. This thermal expansion value is reduced by the aluminum coating but is still much larger than the typical thermal expansion of carbon-fiber reinforced polymer (CFRP) truss components, making 1\% a conservative estimate of differential thermal expansion. To prevent compressive strains and film sagging, the film must be pretensioned to $\sim$1\% tensile strain, which, for even the thinnest unperforated bulk polyimide film, requires forces of several hundred N/m. To reduce this required pretension and impart fabric-like stretchability, we lower the tensile stiffness of the film by perforating it with a kirigami pattern.

Kirigami metamaterials are a class of metamaterials where a desired material response, including but not limited to mechanical and electromagnetic, is controlled by introducing cuts of specific shapes, lengths, and orientations into the bulk material \cite{Jin2024}. Precisely engineered perforations can induce properties not found in the original bulk material, such as auxetic behavior \cite{Rafsanjani2017} and high flexibility or stretchability \cite{Bertoldi2017, An2020}. However, while these cuts improve mechanical compliance, they also reduce electromagnetic reflectance by disrupting material continuity. Therefore, a careful analysis of the tradeoff between mechanical and electromagnetic properties is required for such kirigami film reflectors.

\begin{figure} [h]
\centering
\includegraphics[width=0.95\linewidth]{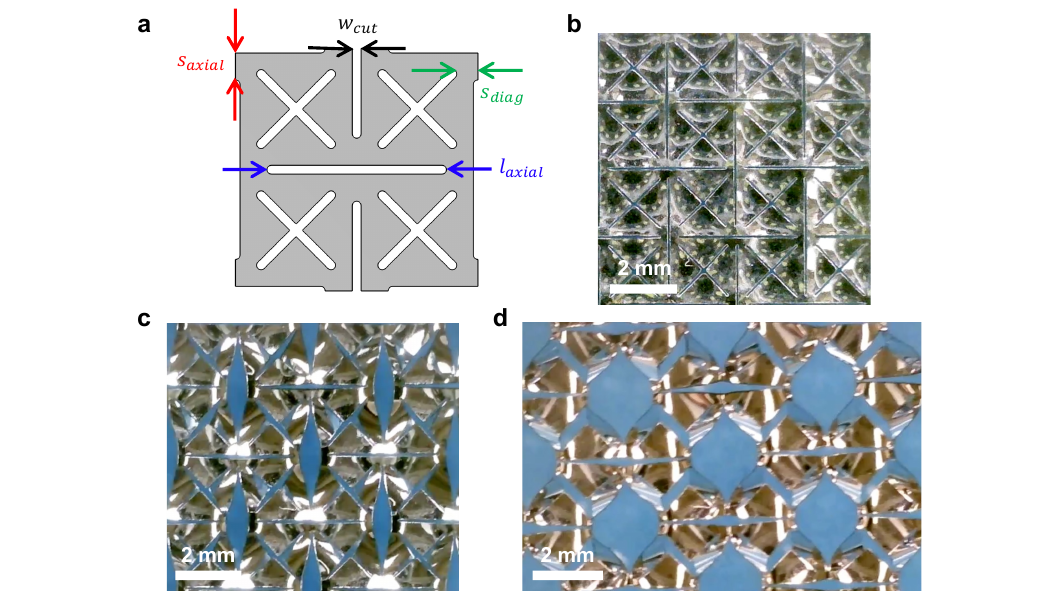}
\caption{(a) A schematic diagram of the unit cell geometry; (b)--(d) photographs of the close-up view of the perforated 7.8~\textmu m thick aluminized polyimide film with $l_{\text{axial}} =$ 4~mm and $s_{\text{axial}} = s_{\text{diag}} = s =$ 0.2~mm (b) in the undeformed state, (c) under $\sim$15\% uniaxial strain forming buckled and sheared squares, and (d) $\sim$45\% uniaxial strain forming larger voids.}
\label{fig:1}
\end{figure}

This work is organized as follows. Sections~\ref{sec:geometry} and~\ref{sec:parametric} introduce the studied kirigami pattern and the simulated trends in how its geometric parameters affect the effective Poisson's ratio, pretension at 1\% strain, and 10~GHz reflectance. Section~\ref{sec:tension} reports the required pretension for the selected designs obtained from their measured and simulated nonlinear tensile responses. Section~\ref{sec:reflectance} presents the power reflectance measurement results for the selected geometries at the target 10~GHz frequency and across 2--12~GHz frequency range, which support the numerical predictions. In section~\ref{sec:summary}, we summarize the overall performance of the tested kirigami reflectors, compare them to the commercially available metallic meshes and those reported in the literature, and outline directions for future work.

\section{Geometry}
\label{sec:geometry}

The presented kirigami film is a periodic structure that consists of a two-dimensional array of unit cells with axial and diagonal cuts (Fig.~\ref{fig:1}a). Axial cuts are perpendicular to each other and located at a fixed distance, separately forming so-called rotating squares \cite{Rafsanjani2017}. At the center of each square, there are two intersecting diagonal cuts placed at another fixed distance from the axial cuts. The novelty of the design lies in these diagonal perforations, which further reduce the tensile stiffness of the film by allowing all squares not only to rotate but also to shear or buckle under stretching in any direction (Fig.~\ref{fig:1}b--d). The unit cell has four independent parameters: the length of the axial cut, $l_{\text{axial}}$, the spacing between two axial cuts, $s_{\text{axial}}$, the spacing between diagonal and axial cuts, $s_{\text{diag}}$, and the width of the cut, $w_{\text{cut}}$.

\section{Finite element parametric studies}
\label{sec:parametric}

Before fabricating the films, we performed a series of parametric finite element simulations in COMSOL Multiphysics to investigate how the unit cell geometry and polyimide thickness affect the pretension at 1\% strain and reflectance at 10~GHz. In the parametric studies, the cut width was fixed at $w_{\text{cut}} =$ 0.1~mm, the polyimide layer thickness was varied from 7.6 to 25.4~\textmu m, and the spacings were varied from 0.1 mm $\leq s_{\text{axial}} = s_{\text{diag}} = s \leq$ 0.3~mm, above which the designs became excessively stiff. The axial cut length was varied from 2~mm $\leq l_{\text{axial}} \leq$ 7~mm. The lower bound was chosen to exclude overly stiff structures, while the upper bound remained well below the target wavelength (3~cm) to avoid diffraction effects. 

The reflectance of a perforated polyimide film coated with 100~nm of aluminum on one side was modeled in the frequency domain at 10~GHz for \textit{s}- and \textit{p}-polarized waves using the wave optics module. A periodic unit cell was modeled inside a rectangular vacuum domain with Floquet periodic boundary conditions applied to the lateral sides. A normally incident polarized wave was sent from a periodic port on the aluminized surface, and a transmitted wave was absorbed by another port located below the unit cell. Since the thicknesses of the polyimide and aluminum layers are orders of magnitude smaller than the unit cell dimensions and the wavelength, the film was defined as a two-dimensional layered material using a layered transition boundary condition. 

In the electromagnetic simulations, the materials were defined by their complex refractive index. The complex refractive index of aluminum was computed using the Drude model assuming plasma frequency of $\omega_{p}=11.9 \times 10^{4}~\mathrm{cm^{-1}}$ and damping frequency of $\omega_{\tau}=6.6 \times 10^{2}~\mathrm{cm^{-1}}$ \cite{Ordal1985}. These parameters correspond to the aluminum DC resistivity of 2.74 ~$\mu\Omega\cdot\text{cm}$ \cite{Ordal1985} and resulted in the calculated complex refractive index at 10~GHz of $\bar{n}=5669.7+i5672.6$. The real and imaginary parts of the refractive index of polyimide were set to 1.8 and 0, respectively, as reported in \cite{Rovensky2018}. We verified the simulation setup by computing reflectance of an unperforated polyimide aluminized on one and two sides with varying aluminum layer thicknesses under normal and oblique incidence using the transfer matrix method \cite{tmm1}, and the results agreed well with the finite element simulations for both polarizations.

Static tensile simulations for perforated polyimide films were conducted in the shell interface using the structural mechanics module. The 100~nm aluminum layer was neglected at this stage because its thickness was negligible compared to the polyimide thickness and was not expected to significantly affect the pretension trends. However, it was later taken into account in the simulations of the promising geometries (see Section~\ref{sec:tension}). The bulk polyimide was modeled as an isotropic material with Young's modulus of $E_{\text{PI}} =$ 2.76~GPa and Poisson's ratio of $\nu_{\text{PI}} =$ 0.34 \cite{DuPont2022}. In these simulations, a linear elastic response was assumed under 1\% tensile strain. Although buckling can occur under applied tension (see Section~\ref{sec:tension}), for thin kirigami sheets, the tensile forces associated with in-plane deformations are typically greater than those for post-buckling out-of-plane deformations. Therefore, this assumption is a conservative estimate for pretension that enabled a more efficient parametric search for soft perforated films. We modeled an infinitely large film as a periodic finite-size strip fixed at one end, with prescribed displacements at the other end, and symmetry boundary conditions applied on the lateral edges. To ensure that boundary effects were negligible, simulations were run for the strips of varying lengths until the reaction force converged with respect to length, which occurred at around sixteen unit cells.

\begin{figure}[H]
    \centering
    \includegraphics[width=1\linewidth]{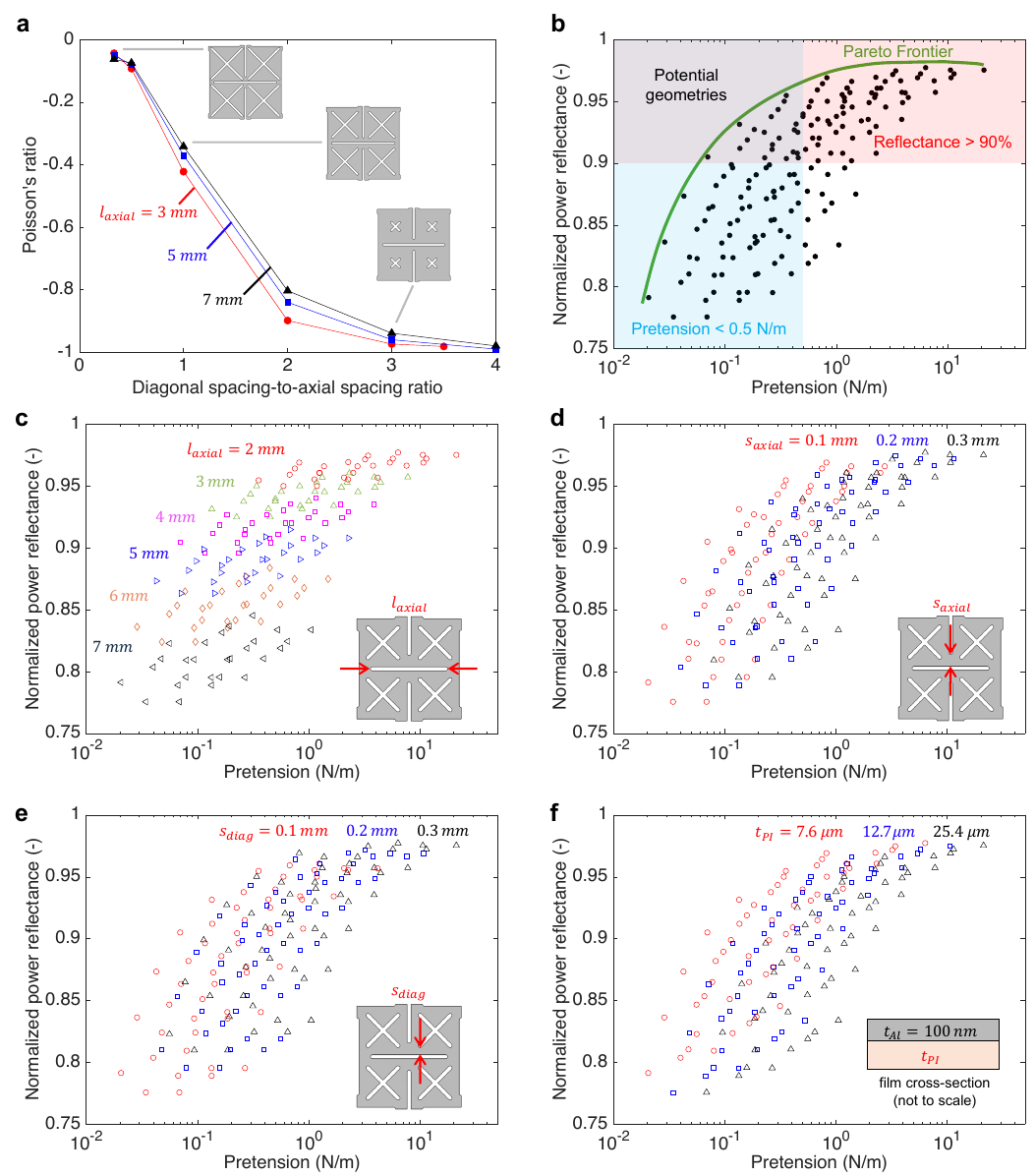}
    \caption{Summary of parametric studies conducted in COMSOL: (a) effect of $s_{\text{diag}}/s_{\text{axial}}$ on the effective Poisson's ratio with a fixed $w_{\text{cut}} =$ 0.1~mm, $s_{\text{axial}} =$ 0.2~mm, and $t_{\text{PI}} =$ 25.4~\textmu m; (b) geometric design space segmented by reflectance and pretension requirements highlighting the Pareto frontier and the region with promising geometries; (c)--(f) color-coded effect of (c) $l_{\text{axial}}$, (d) $s_{\text{axial}}$, (e) $s_{\text{diag}}$, and (f) $t_{\text{PI}}$ on the pretension at 1\% strain and reflectance at 10~GHz.}
    \label{fig:2}
\end{figure}

Additionally, we conducted tensile simulations in the linear elastic regime for finite-size perforated plates under uniaxial tension along axial cuts to determine whether the studied pattern demonstrates auxetic behavior similar to that of the rotating squares geometry. They show that the effective Poisson's ratio of the perforated film can be tuned from $-1$ to nearly 0 by adjusting the ratio of diagonal spacing to axial spacing, $s_{\text{diag}} / s_{\text{axial}}$ (Fig.~\ref{fig:2}a). When $s_{\text{diag}} << s_{\text{axial}}$, shearing of individual squares dominates rotation, resulting in a near-zero Poisson's ratio. Simulations with a fixed axial spacing of 0.2~mm and axial cut lengths of 3~mm, 5~mm, and 7~mm show that Poisson's ratio is near-zero when the diagonal-to-axial spacing ratio is below 0.3. When $s_{\text{diag}} / s_{\text{axial}} > 3.5$, square rotation dominates and produces a strongly auxetic response with a Poisson's ratio approaching $-1$, similarly to the rotating squares geometry \cite{Rafsanjani2017}. When $s_{\text{diag}} = s_{\text{axial}}$, the Poisson's ratio ranges from approximately $-0.3$ to $-0.4$. In the linear elastic regime, the adjustment of the effective Poisson's ratio could allow control over coupling between axial and transverse strains and, as a result, over the in-plane deformations, which could help achieve lower stress concentrations during small-strain thermal cycling.

The pretension and reflectance values obtained from the parametric finite element simulations are summarized in Fig.~\ref{fig:2}b--f, where each point represents a simulated design with a specific axial cut length, axial spacing, diagonal spacing, and polyimide thickness. The designs are positioned according to their required pretension on the $x$-axis and reflectance on the $y$-axis. The results show a tradeoff between pretension and reflectance. Increasing the axial cut length reduces both pretension and reflectance (Fig.~\ref{fig:2}c). Decreasing the axial and diagonal spacings reduces both pretension and reflectance (Fig.~\ref{fig:2}d--e). Trends also show that the axial cut length is the dominant parameter influencing reflectance. Finally, decreasing the polyimide thickness leads to lower pretension with negligible impact on reflectance (Fig.~\ref{fig:2}f).

In this work, we focus on the designs with the power reflectance above 90\% and required pretension below 0.5~N/m, which is an order of magnitude lower than typical pretension requirements for metallic meshes. Based on these criteria, the design space is partitioned as shown in Fig.~\ref{fig:2}b. The plot also highlights a Pareto frontier curve, which includes geometries where one metric (reflectance or pretension) cannot be improved without compromising the other. Designs in the upper-right region of the Pareto frontier show the highest reflectance but require the highest pretension, while those on the lower left are the softest but least reflective. The  optimal geometries in the top left quadrant of the plot generally have axial cuts up to 4~mm, minimal spacings, and the thinnest polyimide layer. Some examples of such optimal geometries were fabricated and studied experimentally as described in the following sections. 

\section{Tensile response of selected designs}
\label{sec:tension}

Guided by the results of the parametric simulations, we selected four designs for experimental tensile testing. Two of them had spacings of $s_{\text{axial}}=s_{\text{diag}}=s=0.2~\text{mm}$ and axial cut length of $l_{\text{axial}}=3~\text{mm}$ and 4~mm. The other two had smaller spacings of around $s_{\text{axial}}=s_{\text{diag}}=s=0.08~\text{mm}$ and axial cut length of $l_{\text{axial}}=3~\text{mm}$ and 3.5~mm. The samples were laser-cut from the bulk 7.6 $\mu\text{m}$ thick polyimide film metalized with 100~nm aluminum on both sides (bulk uncut film supplied by Sheldahl Inc. To quantify pretension, we took the reaction force acting in the direction of stretching on the edge where the extension was prescribed and divided it by the width of the edge.

Linear buckling and post-buckling analysis were conducted in COMSOL for the selected geometries using material properties of the base film. The perforated films were defined as layered materials in the shell interface of the composite materials module with the specified thicknesses and order of aluminum and polyimide layers. The structures were modeled as a periodic strip, fixed at one end, with a prescribed displacement applied at the opposite end, and symmetric or periodic boundary conditions imposed on the lateral sides. To account for the effect of cut orientation within the unit cells, we conducted simulations for at least two directions of uniaxial stretching (axial and diagonal). After the linear buckling analysis, the eigenmode whose shape most closely resembled the experimentally observed periodic deformation was introduced as the initial geometric imperfection for the postbuckling study.

\begin{figure}[H]
    \centering
    \includegraphics[width=1\linewidth]{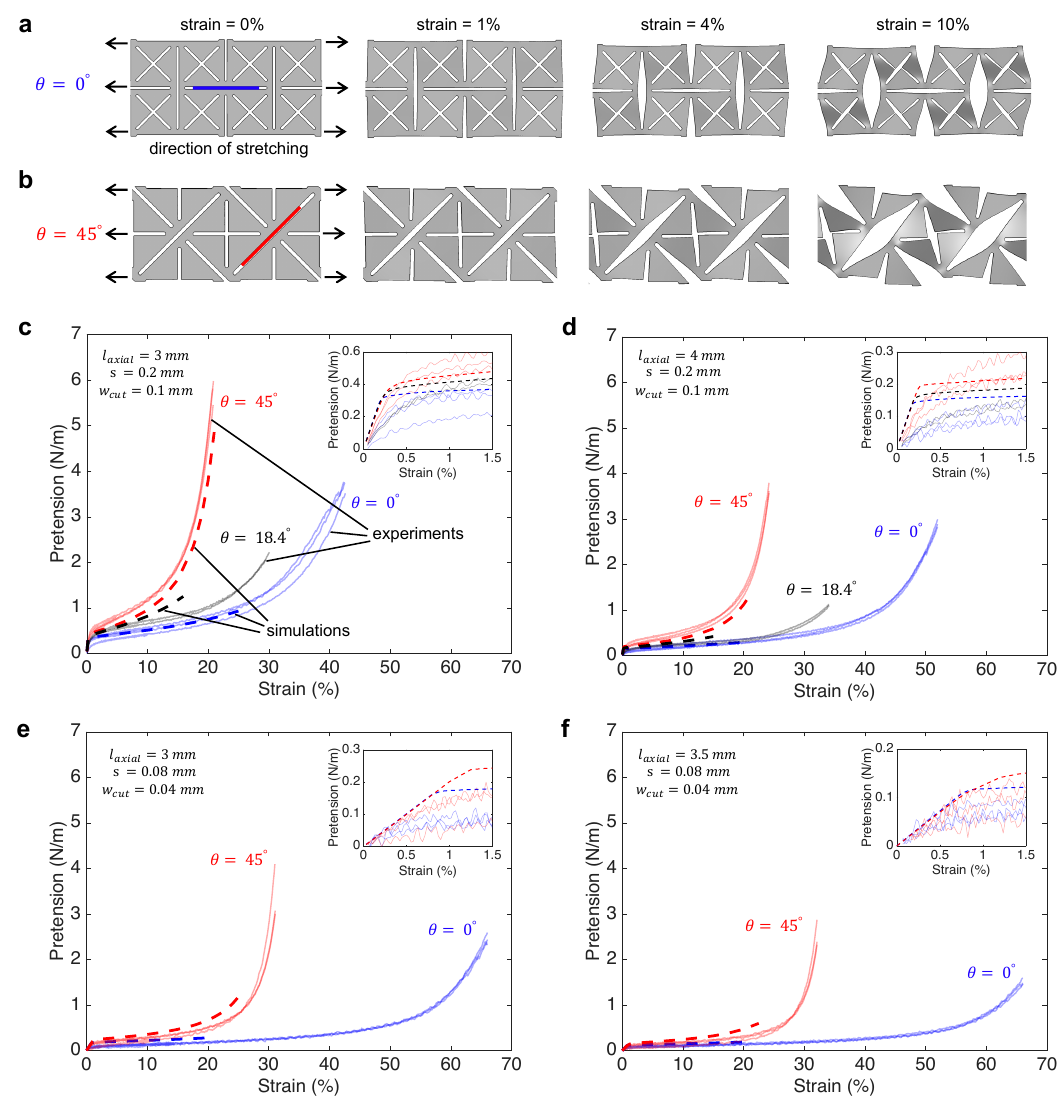}
    \caption{(a)--(b) Fragments of the simulated deformations of the periodic strip with $l_{\text{axial}} =$ 3~\text{mm} and $s =$ 0.2~\text{mm} at different strains under (a) $\theta =$ 0\textdegree{} (blue) and (b) $\theta =$ 45\textdegree{} (red) stretching, where $\theta$ is the angle between the axial cuts and the stretching direction; (c)--(f) results of tensile tests and simulations for selected geometries stretched at $\theta =$ 0\textdegree{} (blue), $\theta =$ 45\textdegree{} (red), and $\theta =$ 18.4\textdegree{} (black) with: (c) $l_{\text{axial}} =$ 3~\text{mm}, $s =$ 0.2~\text{mm}, $w_{\text{cut}} =$ 0.1~\text{mm}, (d) $l_{\text{axial}} =$ 4~\text{mm}, $s =$ 0.2~\text{mm}, $w_{\text{cut}} =$ 0.1~\text{mm}, (e) $l_{\text{axial}} =$ 3~\text{mm}, $s =$ 0.08~\text{mm}, $w_{\text{cut}} =$ 0.04~\text{mm}, and (f) $l_{\text{axial}} =$ 3.5~\text{mm}, $s =$ 0.08~\text{mm}, $w_{\text{cut}} =$ 0.04~\text{mm} with insets enlarging the required pretension at 1\% strain.}
    \label{fig:3}
\end{figure}

Tensile simulations show that the post-buckling pretension of the perforated films varies significantly with the stretching direction as strain increases. (Fig.~\ref{fig:3}). The softest response occurred when the film was stretched along the axial cuts ($\theta =$ 0\textdegree{}) (Fig.~\ref{fig:3}a), while the stiffest response occurred when the film was stretched along the diagonal cuts ($\theta =$ 45\textdegree{}) (Fig.~\ref{fig:3}b). An intermediate orientation ($\theta=$ 18.4\textdegree{}) yielded pretension values between these two extremes. Unphysical self-contact within the deformed film was observed for $\theta =$ 0\textdegree{} at strains above 20--25\%, when the tips of adjacent triangular segments began to overlap due to stretching. Therefore, simulation results are reported up to these strain levels.

Experimental tensile tests showed good agreement with finite element simulations (Fig.~\ref{fig:3}c--f). Samples were tested using a 2.5~N load cell on an Instron 5564 materials testing system. The tested specimens were 9--15 unit cells wide and 24--35 unit cells long. Two to three nominally identical samples were tested for each orientation. Measurements for nominally identical samples were repeatable, with slight variations likely arising from geometric, fabrication, or clamping imperfections. Depending on the stretching direction, the experimentally measured pretension for the 3~mm cut design with 0.2~mm spacings ranged from 0.19 to 0.57~N/m, while for the 4~mm cut design, it decreased to 0.09 to 0.27~N/m (all at 1\% strain). Samples with 0.08~mm spacings showed the smallest pretension, from 0.07~N/m to 0.15~N/m at 1\% strain.  

It is worth noting that pretension remained below 5~N/m in most tested films even at high strains. For example, for the 3~mm sample with 0.2~mm spacing, the pretension reached 5~N/m at 20\% strain under 45\textdegree{} stretching but remained as low as 0.8~N/m at 20\% strain under 0\textdegree{} stretching direction. For the 4~mm sample, pretension at 20\% strain was between 0.37~N/m under 0\textdegree{} and 1.5~N/m under 45\textdegree{} stretching. Films with 0.08~mm spacings showed the lowest pretension, requiring only 0.2--0.5~N/m at 20\% strain. Overall, samples with 0.2~mm spacings were able to reach strains up to 20--25\% in the 45\textdegree{} direction and 40--50\% strains in the 0\textdegree{} direction before breaking, while 0.08~mm spacings slightly extended these limits to around 30\% and 65\%.

\section{RF reflectance of selected designs}
\label{sec:reflectance}

\begin{figure}[H]
    \centering
    \includegraphics[width=1\linewidth]{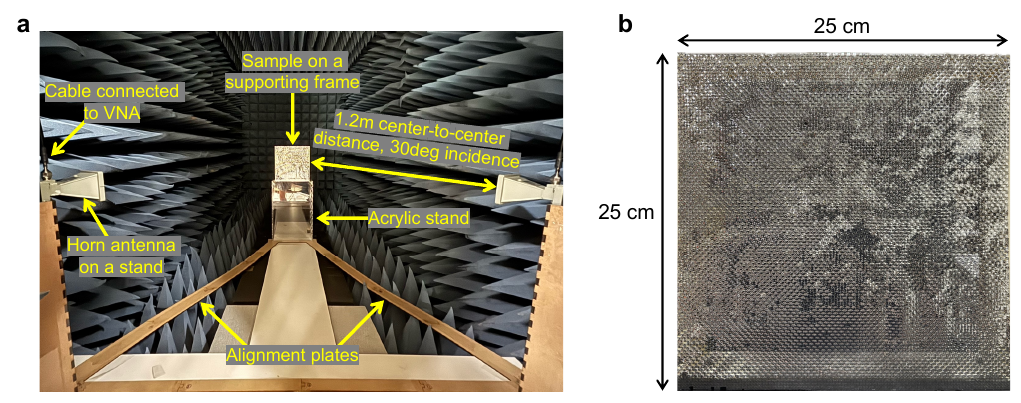}
    \caption {The photographs of (a) RF reflectance measurement setup and (b) one of the 25~cm~$\times$~25~cm tested samples with $l_{\text{axial}} =$ 4~mm and $s =$ 0.2~mm (4.5~mm unit cell period) under around 4\% uniaxial strain mounted on a supporting frame.}
    \label{fig:4}
\end{figure}

We conducted free-space reflectance measurements for selected designs in an anechoic chamber using a two-horn antenna setup at a 30\textdegree{} incidence angle (with respect to the reflector normal) for \textit{s}- and \textit{p}-polarizations (Fig.~\ref{fig:4}a). Two identical horn antennas (Steatite QWH-SL-2-18-N-SG-R) were used, with a 2--18~GHz bandwidth and a 11.8~cm~$\times$~8.5~cm rectangular aperture. Power reflectance was measured over the broadband 2--12~GHz range and at the target 10~GHz frequency by averaging the reflectance spectra from 9.75 to 10.25~GHz. The tested material and horn antennas were positioned 1.2~m from each other, forming an equilateral triangle with horn antenna apertures facing the sample. This distance was chosen to fit the setup within the anechoic chamber and to ensure far-field operation across the entire 2--12~GHz range, which corresponded to roughly 1.1 to 6.5 times the Fraunhofer distance. The position and orientation of the horn antennas and the sample were manually adjusted using the custom-made alignment plates. The tested films, made from the same material used in the mechanical tests, were mounted on 3D-printed polylactide (PLA) frames and suspended above the floor using a clear, RF-transparent acrylic box.

S-parameters were recorded using a vector network analyzer (VNA, Keysight E5063A ENA) over a selected frequency range. Before measurements, the VNA was calibrated using an electronic calibration module to remove cable-dependent effects. The magnitude of the S\textsubscript{21} (or, equivalently, S\textsubscript{12}) was used to obtain the power reflectance. Reflectance measured from each perforated sample was normalized to the reflectance measured from the solid 1/8~in thick aluminum plate (which is commonly used as a reference material, e.g., in \cite{alumX}) or from an unperforated aluminized polyimide film of the same size as the perforated film sample. We verified the high reflectivity of an unperforated film by comparing it to that of the aluminum plate, showing similar
reflectance within our measurement error. Additionally, the 10~GHz power reflectance at \textit{s}-polarization was measured for three reference cases: the empty anechoic chamber, the acrylic stand alone (without a sample), and the supporting frame on top of the acrylic stand (also without a sample). The corresponding reflectance values were 0.02\%, 0.29\%, and 1.74\% respectively. These unwanted reflections were relatively low and within our sample-to-sample measurement error.

We conducted 2--12~GHz broadband reflectance measurements for three selected designs to understand how reflectance changes with frequency for different geometric parameters. We selected two geometries expected to have high reflectance around 10~GHz and one geometry expected to reflect poorly for comparison. The promising samples had 3~mm and 4~mm axial cuts with 0.2~mm spacings (3.5~mm and 4.5~mm periods, respectively, both 25~cm~$\times$~25~cm in size) (Fig.~\ref{fig:4}b), while the intended poor reflector had 9.3~mm cuts with 0.3~mm spacings (10~mm period, 20.5~cm~$\times$~24~cm). The reflectance was measured at both polarizations for perforated films in unstretched and stretched configurations, where stretching was applied uniaxially along the axial cuts. The tested samples were fixed along the bottom edge and, during stretching, any excess material was bent over so that the reflective area facing the antennas remained constant.

\begin{figure}[H]
    \centering
    \includegraphics[width=1\linewidth]{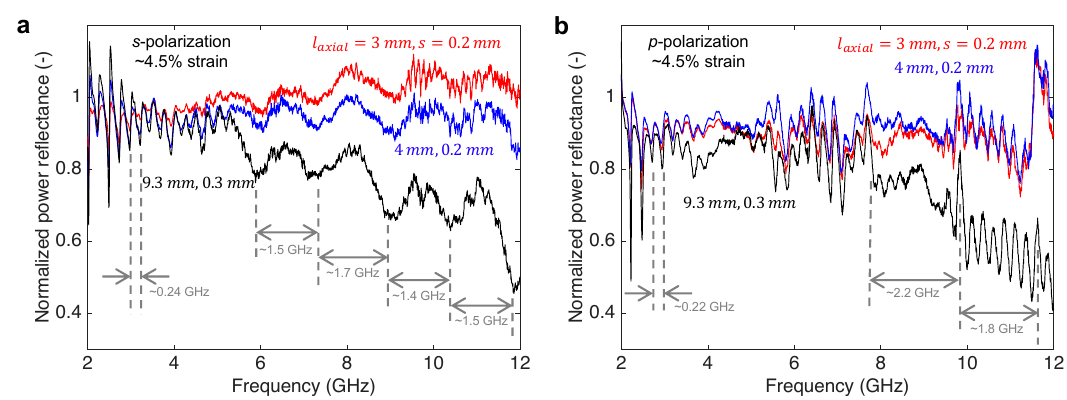}
    \caption {Example of the raw (unsmoothed) normalized power reflectance over 2--12~GHz range for tested perforated films with $l_{\text{axial}} =$ 3~\text{mm}, $s =$ 0.2~\text{mm} (red), $l_{\text{axial}} =$ 4~\text{mm}, $s =$ 0.2~\text{mm} (blue), and $l_{\text{axial}} =$ 9.3~\text{mm}, $s= $ 0.3~\text{mm} (black) under a 30\textdegree{} incidence angle containing similar periodic ripples with the broadest spacings of around (a) 1.5~GHz for \textit{s}-polarization and (b) 2~GHz for \textit{p}-polarization. Here, the reflectance of the 25~cm~$\times$~25~cm samples with 3~mm and 4~mm cuts was normalized with respect to the reflectance of the aluminum plate and the reflectance of the 20.5~cm~$\times$~24~cm sample with 9.3~mm cuts was normalized with respect to the reflectance of the unperforated film of the same size.}
    \label{fig:5}
\end{figure}

The normalized reflectance spectra over the 2--12~GHz range contained ripples with distinct frequency spacings (Fig.~\ref{fig:5}). These ripple patterns varied with polarization but not between samples, indicating that they likely originated from diffraction and multipath interference in the experimental setup rather than from the geometry of the perforated films. For the \textit{s}-polarized wave, the dominant frequency spacing was around 1.5~GHz, while for the \textit{p}-polarized wave it was around 2~GHz, corresponding to the path differences of roughly 20~cm and 15~cm, respectively. Since the polarization was set by the orientation of the horn aperture, the beam shape at the sample plane was different for each case. This could produce distinct diffraction effects, likely due to the acrylic stand, and may explain why the observed path differences scale with the dimensions of the horn aperture in the relevant direction (11.8 and 8.5~cm). In addition to these slow ripples, fast ripples with spacing around 220--240~MHz were observed for both polarizations. They correspond to a path difference of about 1.3~m,  similar to the difference in the the direct path between two horn antennas and the total path length (horn antenna 1 -- sample -- horn antenna 2). To provide a clearer interpretation of the spectral trends, we smoothed the raw normalized reflectance spectra for perforated films using a moving average filter with a window of 1.5~GHz and 2~GHz for \textit{s}- and \textit{p}-polarizations, respectively.

\begin{figure}[H]
    \centering
    \includegraphics[width=1\linewidth]{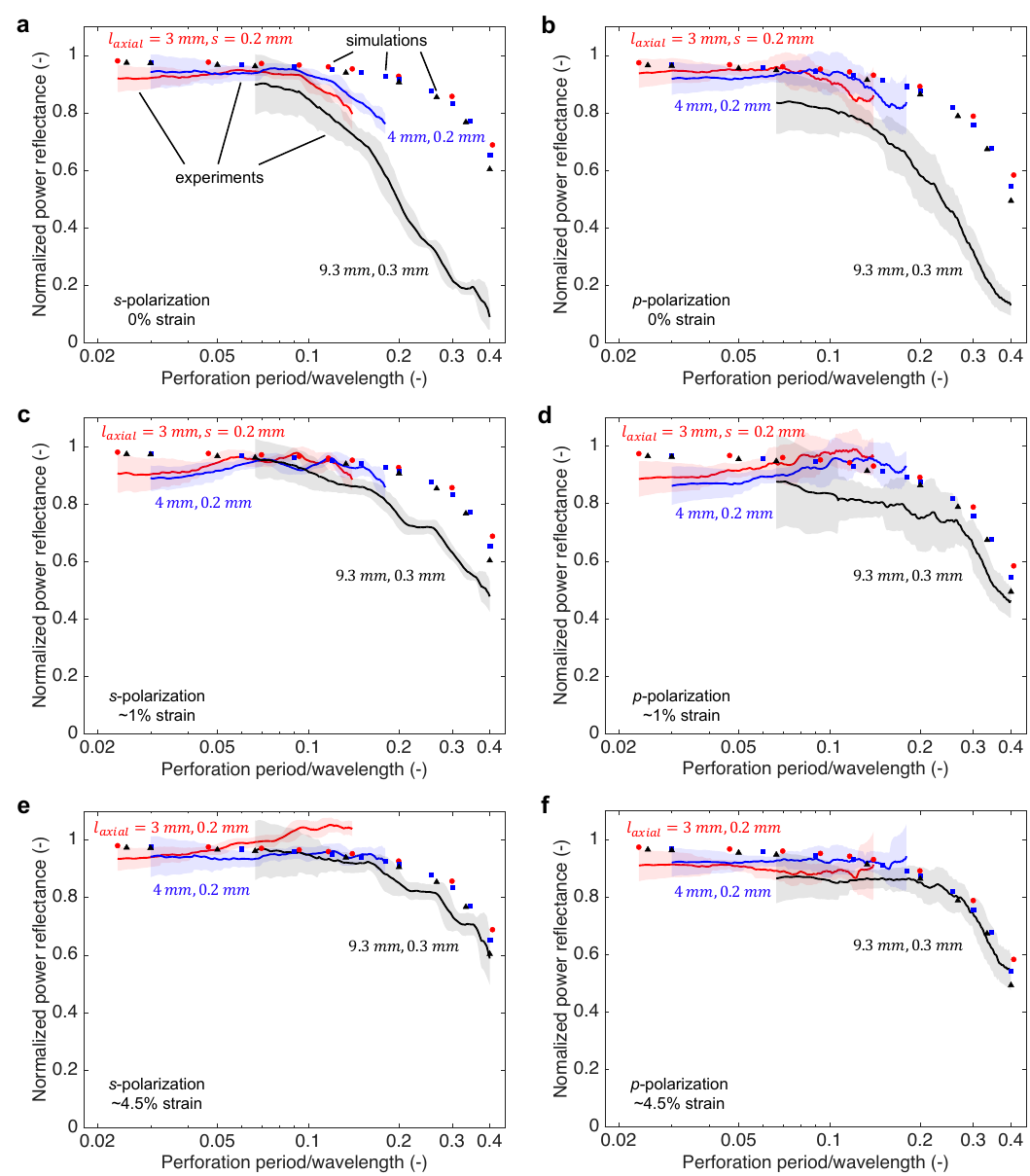}
    \caption {Results of measured and simulated normalized power reflectance for the tested samples with $l_{\text{axial}}=3~\text{mm}$, $s=0.2~\text{mm}$ (red), $l_{\text{axial}}=4~\text{mm}$, $s=0.2~\text{mm}$ (blue), and $l_{\text{axial}}=9.3~\text{mm}$, $s=0.3~\text{mm}$ (black) shown as a function of the perforation period normalized by the wavelength at (a)–(b) 0\%, (c)–(d) $\sim$1\%, and (e)–(f) $\sim$4.5\% strain for \textit{s}- and \textit{p}-polarizations (left and right columns, respectively).}
    \label{fig:6}
\end{figure}

The smoothed normalized power reflectance for each polarization over 2--12~GHz range was plotted for three samples under 0\%, $\sim$1\%, and $\sim$4.5\% strain as a function of the perforation period normalized by the wavelength (Fig.~\ref{fig:6}). The experimentally tested normalized perforation periods ranged from around 0.02 to 0.2 for the samples with 3~mm and 4~mm axial cuts and from around 0.07 to 0.4 for the sample with 9.3~mm axial cuts. To complement experimental results, numerical simulations were performed across the full 0.02--0.4 normalized perforation range: from 2 to 35~GHz for the design with 3~mm cuts (3.5~mm period), from 2 to 26.7~GHz for the design with 4~mm cuts (4.5~mm period), and from 0.75 to 12~GHz for the design with 9.3~mm cuts (10~mm period). 

Before conducting the broadband electromagnetic simulations, we had to update the material properties of the aluminum. The sheet resistances for the experimentally used base unperforated film were 0.55~$\Omega/\square$ and 0.57~$\Omega/\square$ on two sides, which were approximately twice the DC resistance expected for the bulk aluminum resistivity reported by Ordal \cite{Ordal1985}. According to the Drude model, optical resistivity is directly proportional to the damping frequency and inversely proportional to the square of the plasma frequency. We assumed that the optical and DC resistivities are equal and that the plasma frequency remains unchanged since it primarily depends on the electron concentration. In the new set of simulations, conducted for unstretched unit cells under 30\textdegree{} incidence, we therefore increased the damping frequency reported by Ordal by a factor of two to account for the actual resistivity of the experimentally used film.

Simulations provided good predictions of the highest measured reflectance over the 2--12~GHz frequency range (Fig.~\ref{fig:6}). It can be observed that the smaller the unit cell size compared to the wavelength, the higher the reflectance for all three tested unit cell periods. This behavior is expected and has also been reported for wire mesh screens \cite{Casey,liu2013frequency}. The reflectance curves also show that at 0\% strain the experimental samples were generally less reflective than when they are stretched, which was due to wrinkling and/or sagging that occurred when perforated films were oriented vertically (Fig.~\ref{fig:6}a, b). Under 1\% strain, the measured reflectance of 3~mm and 4~mm samples was higher and agreed well with the simulation results (Fig.~\ref{fig:6}c, d). This improvement likely occurred because slight stretching flattened the sample surface, reducing scattering and improving the directionality of the reflected wave. It is possible that even smaller strains could maximize the reflectance of the perforated films. However, such conditions were not tested due to limitations in sample size and manual control of stretching and alignment. The 9.3~mm sample showed lower than predicted reflectance at higher frequencies at around 1\% strain but converged to maximum once stretched to around 4.5\% strain (Fig.~\ref{fig:6}e, f).

\begin{figure}[H]
    \centering
    \includegraphics[width=1\linewidth]{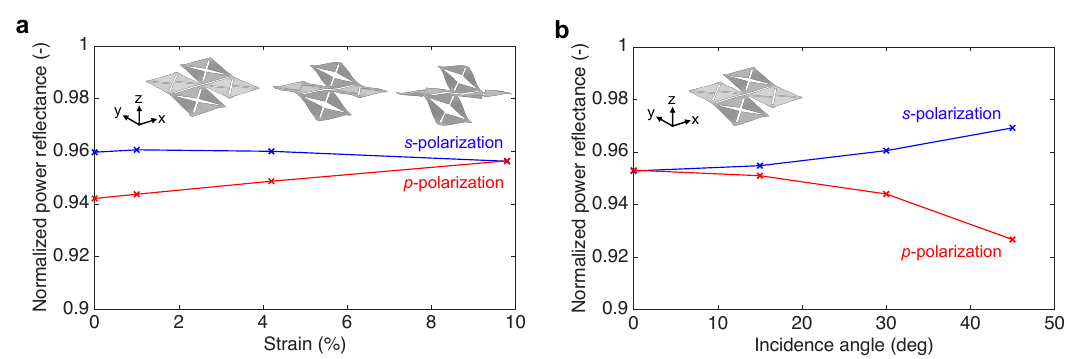}
    \caption {Results of simulated normalized power reflectance for a buckled unit cell with $l_{\text{axial}} =$ 3~\text{mm} and $s =$ 0.2~\text{mm} for \textit{s}- and \textit{p}-polarizations as a function of (a) uniaxial strain under 30\textdegree{} incidence angle and (b) incidence angle for a unit cell under around 1\% uniaxial strain. In both plots, the tension is applied in the $x$-direction, the incident wave is traveling in the negative $z$-direction, and \textit{s}-polarization is defined such that the electric field is oriented in the $y$-direction.}
    \label{fig:7}
\end{figure}

To assess the impact of buckling, we simulated 10~GHz reflectance for a periodic unit cell subjected to uniaxial tension up to 10\% strain. The shapes of the deformed unit cell were first obtained from the post-buckling simulations of a periodic unit cell with 3~mm axial cuts and 0.2~mm spacings. Then, their deformed meshes were imported into the electromagnetic simulation setup. The critical strain of the unit cell was around 0.2\%, beyond which it buckled forming out-of-plane deflections and small voids. However, the reflectance remained almost unchanged, between 94\% and 96\% depending on polarization (Fig.~\ref{fig:7}a), since these features were deeply subwavelength. Reflectance also slightly increased with incidence angle for \textit{s}-polarized waves and decreased for \textit{p}-polarized waves (Fig.~\ref{fig:7}b). Based on these results, we assumed that the reflectance of the samples measured in the stretched configuration could be compared to the reflectance of the simulated undeformed unit cells and that the measured oblique reflectance could still provide a reasonable estimate of the reflectances expected under normal incidence or smaller incidence angles.

\begin{figure}[H]
    \centering
    \includegraphics[width=1\linewidth]{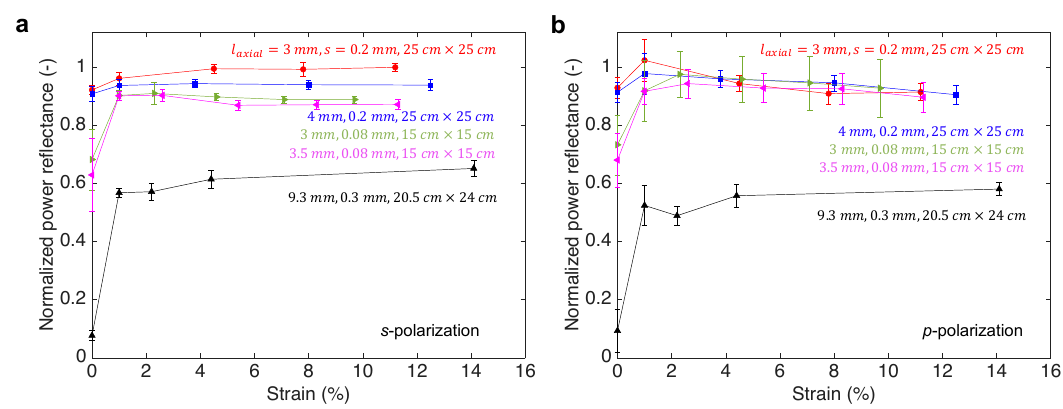}
    \caption {Summary of free-space normalized power reflectance measurements averaged over 9.75--10.25~GHz frequency range for (a) s- and (b) p-polarized waves under a 30\textdegree{} incidence angle for five designs with varying axial cut lengths and spacings in the undeformed and stretched configurations. The mean values were calculated as the average and the error bars represent the standard deviation over the 9.75--10.25 GHz range. The reflectance of the additional  15~cm~$\times$~15~cm samples with 0.08~mm spacings was normalized with respect to the reflectance of the unperforated film of the same size.}
    \label{fig:8}
\end{figure}

The 10~GHz reflectance measurements (averaged over the 9.75--10.25~GHz range) for the three previously tested samples and two additional samples with 3~mm and 3.5~mm axial cuts and 0.08~mm spacings, both 15~cm~$\times$~15~cm in size, are summarized in Fig.~\ref{fig:8}. Each perforated film was tested in unstretched and moderately stretched configurations up to 10--14\% strain. Due to manual setup adjustments and sample placement, each measurement was repeated three times, with the mean values reported as the average and the error bars representing standard deviations.

For \textit{s}-polarization (Fig.~\ref{fig:8}a), the 3~mm design with 0.2~mm spacings showed the highest power reflectance, which increased from 92$\pm$2\% at 0\% strain to 96$\pm$2\% at 1\%, with a slight further increase at higher strains. The 4~mm design followed a similar trend, rising from 91$\pm$2\% in the unstretched state to 94$\pm$2\% at 1\% strain. The reflectance of 3~mm and 3.5~mm designs with 0.08~mm spacings demonstrated slightly lower power reflectance than those with 0.2~mm spacings in the pretensioned state and significantly lower reflectance in the unstretched case. Their reflectance improved from roughly 65\% at no pretension to 90$\pm$2\% at 1\% tensile strain and was nearly constant for higher strains. The decrease in reflectance observed for 3~mm samples with reduced  0.08~mm spacings is likely due to weakened electrical connectivity between adjacent unit cells. As expected, the 9.3~mm structure showed the lowest reflectance, increasing sharply from roughly 8\% in the unstretched state to 57\% at 1\% tensile strain. 

Similar trends and mean reflectance magnitudes are observed for \textit{p}-polarization, however, with an increased measurement error (Fig.~\ref{fig:8}b). The 3~mm and 4~mm samples with 0.2~mm spacings showed moderate increases in reflectance, from about 93\% in the unstretched state to 98--102$\pm$7\% at 1\% tensile strain, followed by a slight decrease at higher strains. For 3~mm and 3.5~mm samples with 0.08~mm spacings, it went up from around 70\% at no pretension to 92$\pm$11\% and 92$\pm$5\% at 1\% tensile strain, respectively. Reflectance of the 9.3~mm design increased significantly from around 10\% to 53\% at 1\% tensile strain and further to nearly 60\% under 14\% tensile strain.

Overall, the narrowband reflectance measurements centered at 10~GHz showed good agreement with the trends observed in the broadband measurements for unstretched and moderately stretched samples (Fig.~\ref{fig:8}). Designs with 3 and 4~mm cuts and 0.2~mm spacings showed power reflectance above 90\% for both polarizations even without pretension. All tests, however, consistently demonstrated that the reflectance of selected promising films can improve under stretching, with a strain of about 1\% being sufficient to achieve high and stable reflectance within our measurement error.

\section{Summary and outlook}
\label{sec:summary}

\begin{figure} [h]
\centering
\includegraphics[width=0.58\linewidth]{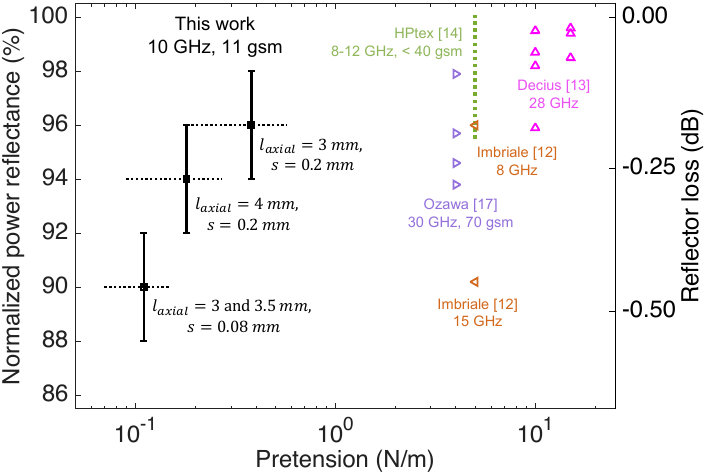}
\caption{Comparison of the performance of the proposed in this work kirigami film reflectors with a commercial metallic mesh \cite{hptex} and meshes reported in the literature \cite{Imbriale1991, Ozawa2023, Decius2023} at the specified testing frequencies and with areal densities in grams per square meter (gsm) where reported. Plotted reflectance values for selected kirigami films are weighted averages of the measured power reflectances for \textit{s}- and \textit{p}-polarizations at 1\% strain under 30\textdegree{} incidence over 9.75--10.25~GHz with error bars denoting weighted standard deviations. The dashed horizontal lines show the range of measured required pretensions with left and right bounds corresponding to $\theta =$ 0\textdegree{} and 45\textdegree{} stretching orientations, respectively. References \cite{hptex, Imbriale1991} did not specify pretension; therefore, a nominal value of 5~N/m was assumed. Reflectance values from \cite{Decius2023} were measured at~28 GHz under 40\textdegree{} incidence and pretensions of 10 and 15~N/m, while \cite{Ozawa2023} reported measurements at 30~GHz under normal incidence and mentioned an intended usage at 4~N/m. Note: a vertical line for \cite{hptex} represents the minimum reported power reflectance with maximum assumed to be 100\%.}
\label{fig:9}
\end{figure}

The performance of the selected tested geometries is summarized in Fig.~\ref{fig:9} and compared with a commercial metallic mesh and some designs reported in the literature. At the target 1\% strain, the required pretension for the kirigami films is nearly an order of magnitude lower than the metal meshes, demonstrating significant improvement in mechanical performance and enabling lighter and more stowable truss architectures.

While the mean power reflectance is slightly lower than the power reflectance of metallic meshes at higher frequencies, there is a potential to achieve comparable performance through targeted design optimization. One approach to increase the power reflectance is to reduce the cut length and/or increase the spacing by partially compromising the pretension. Another strategy involves increasing the thickness of the aluminum layer while using a thinner polyimide substrate to maintain low stiffness. Alternatively, employing a more conductive aluminum coating or metals with lower sheet resistance could also further enhance reflectance.

The ability of the studied kirigami films to withstand the construction and deployment processes is also yet to be verified, as these operations are far more complex than the controlled uniaxial tension applied in the Instron tests. During storage, unfolding from a stowed configuration, or antenna assembly, localized regions of high, multidirectional strain and twisting may develop. Because several-micron-thick perforated films are delicate, extreme strains and twisting must be avoided to prevent tearing. However, the magnitude and direction of such critical deformations can, in principle, be predicted and quantified during ground testing. This information could guide the selection of films with tailored tensile responses and inform their orientation to minimize damage and reduce the risk of crack formation and propagation.

To conclude, we proposed a low-pretension, RF-reflective kirigami thin film as an alternative reflective material for large space-based antennas and presented one possible kirigami design that combines low required pretension, low areal density, and high RF reflectance. As part of future work, alternative geometries and materials could be explored to further enhance performance across a wider range of radio frequency bands. Building on the work in Refs. \cite{Dassanayake2024, dassanayake2025numerical, Suh2024}, additional studies could also be conducted to quantify how much using such low-pretension reflecting films will reduce the mass and stowed volume of the large deployable antenna trusses.

\section*{Acknowledgment}
This work was supported by the DARPA NOM4D program (Grant No. HR001122C0054) directed by Dr. Andrew Detor. We are grateful for the valuable guidance provided by our collaborators at Caltech, Professor Sergio Pellegrino and Dr. Jongeun Suh. We also thank Professor Troy Olsson at the University of Pennsylvania for providing access to the anechoic chamber.


\begin{thebibliography}{33}
\newcommand{\enquote}[1]{``#1''}
\providecommand{\natexlab}[1]{#1}
\providecommand{\url}[1]{\texttt{#1}}
\providecommand{\urlprefix}{URL }
\expandafter\ifx\csname urlstyle\endcsname\relax
  \providecommand{\doi}[1]{\discretionary{}{}{}https://doi.org/#1}\else
  \providecommand{\doi}[1]{\discretionary{}{}{}\urlstyle{rm}\url{https://doi.org/#1}}\fi

\bibitem[{Schenk et~al.(2014)Schenk, Viquerat, Seffen, and Guest}]{Schenk2014}
Schenk, M., Viquerat, A.~D., Seffen, K.~A., and Guest, S.~D., \enquote{Review of Inflatable Booms for Deployable Space Structures: Packing and Rigidization,} \emph{Journal of Spacecraft and Rockets}, Vol.~51, No.~3, 2014, pp. 762--778.
\newblock \doi{10.2514/1.A32598}.

\bibitem[{Miura and Pellegrino(2020)}]{miura2020forms}
Miura, K., and Pellegrino, S., \emph{Forms and concepts for lightweight structures}, Cambridge University Press, 2020.
\newblock \doi{10.1017/9781139048569}.

\bibitem[{Tibert and Pellegrino(2002)}]{Tibert2002}
Tibert, A.~G., and Pellegrino, S., \enquote{Deployable tensegrity reflectors for small satellites,} \emph{Journal of Spacecraft and Rockets}, Vol.~39, 2002, pp. 701--709.
\newblock \doi{10.2514/2.3867}.

\bibitem[{Hansen(1981)}]{Hansen1981}
Hansen, R., \enquote{Fundamental limitations in antennas,} \emph{Proceedings of the IEEE}, Vol.~69, 1981, pp. 170--182.
\newblock \doi{10.1109/PROC.1981.11950}.

\bibitem[{{Northrop Grumman Systems Corporation}(2017)}]{northrop2017}
{Northrop Grumman Systems Corporation}, \enquote{AstroMesh Unfurlable Mesh Antenna,} Technical datasheet, 2017.
\newblock \urlprefix\url{https://cdn.northropgrumman.com/-/media/wp-content/uploads/AstroMesh-DataSheet.pdf/}.

\bibitem[{Thomson(1999)}]{Thomson1999}
Thomson, M., \enquote{The AstroMesh deployable reflector,} \emph{IEEE Antennas and Propagation Society International Symposium. 1999 Digest. Held in conjunction with: USNC/URSI National Radio Science Meeting (Cat. No.99CH37010)}, IEEE, 1999, pp. 1516--1519.
\newblock \doi{10.1109/APS.1999.838231}.

\bibitem[{Powell(1978)}]{POWELL1978}
Powell, R., \enquote{A future for large space antennas,} \emph{7th Communications Satellite Systems Conference}, 1978, p. 588.
\newblock \doi{10.2514/6.1978-588}.

\bibitem[{Lee et~al.(2016)Lee, Backes, Burdick, Pellegrino, Fuller, Hogstrom, Kennedy, Kim, Mukherjee, Seubert, and Wu}]{Lee2016}
Lee, N., Backes, P., Burdick, J., Pellegrino, S., Fuller, C., Hogstrom, K., Kennedy, B., Kim, J., Mukherjee, R., Seubert, C., and Wu, Y.-H., \enquote{Architecture for in-space robotic assembly of a modular space telescope,} \emph{Journal of Astronomical Telescopes, Instruments, and Systems}, Vol.~2, 2016, p. 041207.
\newblock \doi{10.1117/1.JATIS.2.4.041207}.

\bibitem[{Suh et~al.(2024)Suh, Dassanayake, Thomson, and Pellegrino}]{Suh2024}
Suh, J.~E., Dassanayake, S.~P., Thomson, M.~W., and Pellegrino, S., \enquote{Concept for Scalable Mesh Reflector Antennas Assembled in Space,} \emph{AIAA SciTech Forum and Exposition, 2024}, American Institute of Aeronautics and Astronautics Inc, AIAA, 2024.
\newblock \doi{10.2514/6.2024-0823}.

\bibitem[{Dassanayake et~al.(2025)Dassanayake, Suh, and Pellegrino}]{dassanayake2025numerical}
Dassanayake, S.~P., Suh, J.-E., and Pellegrino, S., \enquote{Numerical study of novel concept for in-space assembly of ring-like space structures,} \emph{European Journal of Mechanics-A/Solids}, 2025, p. 105790.
\newblock \doi{10.1016/j.euromechsol.2025.105790}.

\bibitem[{Dassanayake et~al.(2024)Dassanayake, Suh, Thomson, and Pellegrino}]{Dassanayake2024}
Dassanayake, S.~P., Suh, J.~E., Thomson, M.~W., and Pellegrino, S., \enquote{Mass, Volume and Natural Frequency Scaling of Deployable Mesh Reflectors,} \emph{AIAA SciTech Forum and Exposition, 2024}, American Institute of Aeronautics and Astronautics Inc, AIAA, 2024.
\newblock \doi{10.2514/6.2024-2041}.

\bibitem[{Imbriale et~al.(1991)Imbriale, Galindo-Israel, and Rahmat-Samii}]{Imbriale1991}
Imbriale, W.~A., Galindo-Israel, V., and Rahmat-Samii, Y., \enquote{On the reflectivity of complex mesh surfaces (spacecraft reflector antennas),} \emph{IEEE transactions on antennas and propagation}, Vol.~39, 1991, pp. 1352--1365.
\newblock \doi{10.1109/8.99044}.

\bibitem[{Decius et~al.(2023)Decius, Hoeck, Salvini, Marotta, Moseley, Ihle, Rodrigues, Suess, and Angevain}]{Decius2023}
Decius, M., Hoeck, S., Salvini, P., Marotta, E., Moseley, P., Ihle, A., Rodrigues, G., Suess, M., and Angevain, J.-C., \enquote{Advanced RF reflective metal mesh for high frequency deployable reflector antennas,} \emph{Proceedings of 41st ESA Antenna Workshop on Large Deployable Antennas}, 2023.

\bibitem[{{High Performance Textiles GmbH}(2025)}]{hptex}
{High Performance Textiles GmbH}, \enquote{Universal Space Mesh - USM-ATLAS32,} Technical datasheet, 2025.
\newblock \urlprefix\url{https://www.hptex.de/wp-content/uploads/2025/09/HPtex-USM-ATLAS32_Data-Sheet_2025-09.pdf}.

\bibitem[{Hedgepeth and Adams(1983)}]{Hedgepeth1983}
Hedgepeth, J.~M., and Adams, L.~R., \enquote{Design Concepts for Large Reflector Antenna Structures,} Tech. Rep. NASA-CR-3663, NASA, 1983.
\newblock \urlprefix\url{https://ntrs.nasa.gov/citations/19830008513}.

\bibitem[{Santiago-Prowald and Baier(2013)}]{santiago}
Santiago-Prowald, J., and Baier, H., \enquote{Advances in Deployable Structures and Surfaces for Large Apertures in Space,} \emph{CEAS Space Journal}, Vol.~5, No. 3-4, 2013, pp. 89--115.
\newblock \doi{10.1007/s12567-013-0048-3}.

\bibitem[{Ozawa et~al.(2023)Ozawa, Nakamura, Daisuke, Muramatsu, Fujii, and Mori}]{Ozawa2023}
Ozawa, S., Nakamura, K., Daisuke, M., Muramatsu, N., Fujii, H., and Mori, M., \enquote{Electrical and mechanical properties of metal mesh for Ka-band using zirconium copper wire,} \emph{Proceedings of 41st ESA Antenna Workshop on Large Deployable Antennas}, 2023.

\bibitem[{Ji et~al.(2019)Ji, Li, Hu, and Fuchs}]{Ji2019}
Ji, D., Li, T., Hu, W., and Fuchs, H., \enquote{Recent Progress in Aromatic Polyimide Dielectrics for Organic Electronic Devices and Circuits,} \emph{Advanced Materials}, Vol.~31, 2019.
\newblock \doi{10.1002/adma.201806070}.

\bibitem[{Gouzman et~al.(2019)Gouzman, Grossman, Verker, Atar, Bolker, and Eliaz}]{Gouzman2019}
Gouzman, I., Grossman, E., Verker, R., Atar, N., Bolker, A., and Eliaz, N., \enquote{Advances in Polyimide-Based Materials for Space Applications,} \emph{Advanced Materials}, Vol.~31, No.~18, 2019.
\newblock \doi{10.1002/adma.201807738}.

\bibitem[{Mori et~al.(2010)Mori, Sawada, Funase, Morimoto, Endo, Yamamoto, Tsuda, Kawakatsu, Kawaguchi, Miyazaki, and Shirasawa}]{ikaros}
Mori, O., Sawada, H., Funase, R., Morimoto, M., Endo, T., Yamamoto, T., Tsuda, Y., Kawakatsu, Y., Kawaguchi, I., Miyazaki, Y., and Shirasawa, Y., \enquote{First Solar Power Sail Demonstration by IKAROS,} \emph{Transactions of the Japan Society for Aeronautical and Space Sciences, Aerospace Technology Japan}, Vol.~8, No. ists27, 2010, pp. 4--25.
\newblock \doi{10.2322/tastj.8.To_4_25}.

\bibitem[{Katan(2012)}]{katan2012nasa}
Katan, C., \enquote{NASA's Next Solar Sail: Lessons Learned from Nanosail-D2,} \emph{26th Annual AIAA/USU Conference on Small Satellites: Enhancing Global Awareness through Small Satellites}, 2012.
\newblock \urlprefix\url{https://ntrs.nasa.gov/citations/20120015556}.

\bibitem[{{Sheldahl, Inc.}(2020)}]{Sheldahl2020}
{Sheldahl, Inc.}, \enquote{The Red Book,} Technical datasheet, 2020.
\newblock \urlprefix\url{https://www.sheldahl.com}.

\bibitem[{{DuPont}(2022)}]{DuPont2022}
{DuPont}, \enquote{Dupont Kapton Summary of Properties,} Technical datasheet, 2022.
\newblock \urlprefix\url{https://www.dupont.com/content/dam/electronics/amer/us/en/electronics/public/documents/en/EI-10142_Kapton-Summary-of-Properties.pdf}.

\bibitem[{Jin and Yang(2024)}]{Jin2024}
Jin, L., and Yang, S., \enquote{Engineering Kirigami Frameworks Toward Real-World Applications,} \emph{Advanced Materials}, Vol.~36, No.~9, 2024.
\newblock \doi{10.1002/adma.202308560}.

\bibitem[{Rafsanjani and Bertoldi(2017)}]{Rafsanjani2017}
Rafsanjani, A., and Bertoldi, K., \enquote{Buckling-Induced Kirigami,} \emph{Physical Review Letters}, Vol. 118, 2017.
\newblock \doi{10.1103/PhysRevLett.118.084301}.

\bibitem[{Bertoldi et~al.(2017)Bertoldi, Vitelli, Christensen, and Hecke}]{Bertoldi2017}
Bertoldi, K., Vitelli, V., Christensen, J., and Hecke, M.~V., \enquote{Flexible Mechanical Metamaterials,} \emph{Nature Reviews Materials}, Vol.~2, 2017.
\newblock \doi{10.1038/natrevmats.2017.66}.

\bibitem[{An et~al.(2020)An, Domel, Zhou, Rafsanjani, and Bertoldi}]{An2020}
An, N., Domel, A.~G., Zhou, J., Rafsanjani, A., and Bertoldi, K., \enquote{Programmable Hierarchical Kirigami,} \emph{Advanced Functional Materials}, Vol.~30, 2020.
\newblock \doi{10.1002/adfm.201906711}.

\bibitem[{Ordal et~al.(1985)Ordal, Bell, Alexander, Long, and Querry}]{Ordal1985}
Ordal, M.~A., Bell, R.~J., Alexander, R.~W., Long, L.~L., and Querry, M.~R., \enquote{Optical properties of fourteen metals in the infrared and far infrared,} \emph{Applied optics}, Vol.~24, 1985, pp. 4493--4499.
\newblock \doi{https://doi.org/10.1364/AO.24.004493}.

\bibitem[{Rovensky et~al.(2018)Rovensky, Pietrikova, and Lukacs}]{Rovensky2018}
Rovensky, T., Pietrikova, A., and Lukacs, P., \enquote{Dielectric properties' homogeneity of various substrates in GHz area,} \emph{41st International Spring Seminar on Electronics Technology (ISSE)}, 2018.
\newblock \doi{10.1109/ISSE.2018.8443761}.

\bibitem[{Macleod and Macleod(2010)}]{tmm1}
Macleod, H.~A., and Macleod, H.~A., \emph{Thin-film optical filters}, CRC press, 2010.
\newblock \doi{10.1201/9781420073034}.

\bibitem[{Asadchy et~al.(2017)Asadchy, D{\'\i}az-Rubio, Tcvetkova, Kwon, Elsakka, Albooyeh, and Tretyakov}]{alumX}
Asadchy, V., D{\'\i}az-Rubio, A., Tcvetkova, S., Kwon, D.-H., Elsakka, A., Albooyeh, M., and Tretyakov, S., \enquote{Flat engineered multichannel reflectors,} \emph{Physical Review X}, Vol.~7, No.~3, 2017, p. 031046.
\newblock \doi{10.1103/PhysRevX.7.031046}.

\bibitem[{Casey(1988)}]{Casey}
Casey, K., \enquote{Electromagnetic shielding behavior of wire-mesh screens,} \emph{IEEE Transactions on Electromagnetic Compatibility}, Vol.~30, No.~3, 1988, pp. 298--306.
\newblock \doi{10.1109/15.3309}.

\bibitem[{Liu and Tan(2013)}]{liu2013frequency}
Liu, Y., and Tan, J., \enquote{Frequency dependent model of sheet resistance and effect analysis on shielding effectiveness of transparent conductive mesh coatings,} \emph{Progress In Electromagnetics Research}, Vol. 140, 2013, pp. 353--368.
\newblock \doi{10.2528/PIER13050312}.

\end{thebibliography}
\end{document}